\documentclass[12pt,preprint]{aastex}

\shorttitle{Orientation of Planetary Nebulae} \shortauthors{Weidmann
\& D\'{\i}az}

\begin{document}

\title{The Spatial Orientation of Planetary Nebulae\\ Within the Milky~Way}

\author{Walter A. Weidmann}
\affil{Observatorio Astron\'omico de C\'ordoba and CONICET, Laprida
854, 5000 C\'ordoba, Argentina.} \email{walter@oac.uncor.edu}

\and

\author{Rub\'en J. D\'{\i}az}
\affil{Gemini Observatory, AURA, La Serena, Chile.\\
Complejo Astron\'omico El Leoncito, CONICET, Argentina.}
\email{rdiaz@gemini.edu}

\begin{abstract}
  We analyze the spatial orientation of a homogenous sample of
  440 elongated Planetary Nebulae (PNe) in order to determine the
  orientation of their apparent major axis respect to the Milky
  Way plane. We present some important geometrical and statistical
  considerations that have been overlooked by the previous works on the subject.
  The global distribution of galactic position angles (GPA) of PNe is
  quantitatively not very different from a random distribution of
  orientations in the Galaxy.
  Nevertheless we find that there is at least one region
  on the sky, toward  the galactic center, where a weak correlation
  may exist between the orientation of the major axis of some PNe and the Galactic equator,
  with an excess of axes with GPA$\sim 100^{\circ}$.

  Therefore, we confirm that ``extrinsic'' phenomena (i.e., global galactic
  magnetic fields, shell compression from motion relative to the Interstellar
  Medium) do not determine the morphology of PNe on most of the sky,
  with a possible exception towards the galactic center.

\end{abstract}


\section{INTRODUCTION}

Since Charles Messier registered the first Planetary Nebulae
(Dumbell Nebula in Vulpecula) on July 12, 1764 until the present
time, over one thousand of these beautiful objects were found in our
Galaxy and much more in nearby galaxies. It is well known that PNe
have different shapes but, in most cases,    the projection of a
nebula onto the sky has a defined extension axis. Besides, the study
of the orientation of many kind of phenomena such as supernova
remnants (Gaensler 1998), orbits of binary systems (Brazhnikova et
al. 1975), rotation of individual stars (Hensberge et al. 1979), has
been of broad astrophysical interest. This sort of study provides
information about formation, evolution and death of the stars.

However, the orientation of the projected major axis of the PNe was
rarely studied and the results that were found are still
contradictory.

For instance, Shain (1956) and Gurzadyan (1958) made the pioneering
works about the orientation of PNe. They worked with very small
samples from the Curtis catalogue (1917) and their conclusions were
contradictory. Shain found that the angle between the semi-major
axis of PNe and the galactic equator was small for objects with low
latitudes. On the other hand Gurzadyan did not found any correlation
and concluded that the magnetic field could not be influencing the
shape of PNe. At the present time both papers only have a historical
character. The first works to determine spatial orientations of PNe
with a significant number of objects were those of Grinin \& Zvereva
(1968), and Melnick \& Harwitt (1974). These works were more
complete than early ones and both papers concluded that PNe are
aligned with the plane. The hypothesis explaining these non-random
orientations were based in the effect of the ambient magnetic field
and interactions with the interstellar medium.

The works of Phillips (1997) and Corradi, Aznar \& Mampaso (1998;
hereafter CAM98) are the most recent and contradictory  papers about
global PNe orientations. In both cases, the images examined had high
quality and the determination of PA have small errors, but they did
not study the orientation of PNe over a specific region of the sky.
The sample of Phillips was culled from a variety of published images
complemented with broad band survey plates, whereas the sample of
Corradi and coworkers comes mainly from three different narrow band
surveys.

In this paper we visit this topic and make a careful study with new
geometrical considerations. The characterization of any found
correlation in a certain region of the sky would help to disentangle
the role of the galactic magnetic field in determining the
morphologies of these sources. Alternatively, it could be found that
the motion of PNe through the interstellar medium leads to some
compression of their shells, and result in significant correlations
between the apparent outflow axes.

The unrestricted total sample was extracted from 3 different
sources: Digitized Sky Survey version 2, DSS2 (red broad band); the
Macquarie$/AAO/$Strasbourg $H_\alpha$ Planetary Nebulae Catalog,
(MASH; $H_\alpha$ narrow band); and HST archival imagery (filters
F502N and F656N).  After the starting criteria described in Section
2, the initial number of PNe studied in this work resulted in 868,
significantly larger than in previous studies. The analysis is
performed through Sections 3 to 5, the results are presented and
discussed in Sections 6 and 7.

\section{SELECTION OF THE SAMPLE}

The sample was extracted from the Catalogue of Galactic Planetary
Nebulae updated version 2000 (Kohoutek 2001; hereafter CGPN2000)
that includes 1510 true PNe and the MASH survey (Parker et al. 2006)
with 903 objects, 578 of them are classified as true PN.  There is
virtually no overlap between the CGPN2000 and the MASH catalogs,
therefore this last one provided one third of the final sample of
truly elongated PNe. We included a special subsample of the CGPN2000
which is formed by those objects imaged by the Hubble Space
Telescope (HST).

The selection criteria employed to obtain our final sample were:

\begin{enumerate}
\item Objects classified as true PN.

\item Angular size larger than 10$\arcsec$. The experience shows that,
in general, this limit yields spatial resolution sufficient to
determine the orientation of the major axis.  Exception in the
angular size criterium was made for those objects imaged by the HST.

\item $b<20^{\circ}$, to avoid information degradation by projection
effects over high latitudes objects.

\item In case of the MASH survey we only included PNe which at least have one
confirmation spectrum.

\item Objects clearly visible in the DSS2 plates (surface brightness $<$ 17
mag/arcmin$^{2}$), this restriction criterium was applied for all
the surveys contributing to the sample.

\item Elongated shape (PNe with morphology type R  were rejected).
\end{enumerate}

By elongated shape PNe we mean bipolar nebulae or those which show
two lobes defining unambiguously the direction of their polar
outflows, as well as E nebulae with an appreciable ellipticity. At
the cases where the object is near to the DSS resolution we estimate
an empirical limit of major to minor axis length ratio 1.2:1, as
determined from those objects which were in the visual limit of
acceptance.

In order to extend our sample we included PNe with angular size
between 4$\arcsec$ and 10$\arcsec$ inclusively, that had been
observed by HST. Such PNe, 42 in total number and 27 truly
elongated, belong to some of our 4 regions. Due to the variety of
filters the chosen criterium here was to use the images taken
through the F656N or F502N filters whenever possible. To complete
the sample we also included those PNe measured by CAM98 that we
could not measure from the DSS2 plates. After a careful examination
of more than 868 PNe in the initial list (obtained after criteria 1,
2 and 3), we arrived to the final sample that contains 444 truly
elongated PNe distributed in four regions, 174 belonging to the MASH
sample. The remaining objects were planetaries with apparent
circular shape, low surface brightness or very peculiar morphology.

\section{MEASUREMENT OF POSITION ANGLES}

\begin{figure}
\includegraphics{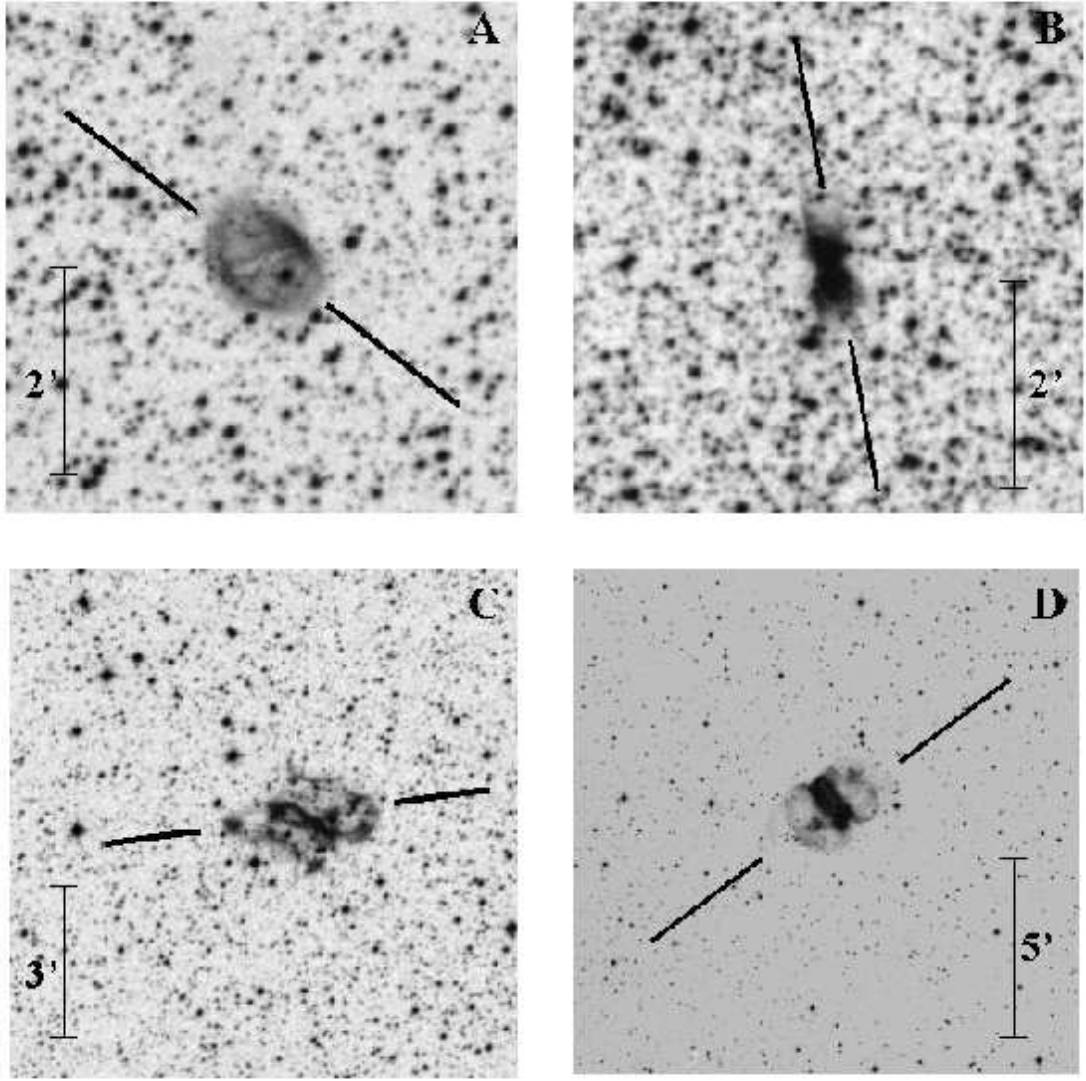}
\vspace{16cm} \caption{Examples of the applied measuring criteria.
The objects are: A)~NGC~4071, B)~Mz3, C)~NGC~5189 and D)~NGC~650.
All the images belong to the DSS, red band. North is up and East to
the left.} \label{fotito1}
\end{figure}

We measured the position angles (PA) of the projected major axis
over the sky, for all the PNe of the final sample. PA are measured
following the usual convention: from the north towards the east.
These angles are measured within the equatorial coordinates system
so we call them EPA. The directions of the elongated axis were
estimated visually by fixing the line that better represents the
long symmetry axis of the PN. The angle that this line forms with
the north was the EPA. For bipolar structures the EPA was measured
with respect to the direction of the outflow of the objects. The
uncertainties of EPA are estimated following the criterium of CAM98:
all measurements were repeated independently by both authors. The
largest difference between both measurements was 8$^{\circ}$. The
measuring criteria are exemplified in Fig.~\ref{fotito1}.

In spite of the fact that all previous works determined the EPA
visually, we tried to obtain those angles through an automated way,
but the problem was that this sort of automated algorithms are still
not powerful enough to deal with low S/N and overcrowded fields with
high galactic background. PNe are usually located in highly
populated Milky Way fields and in many cases do not show structure
complete at the same surface brightness. This patchy structure and
the star images make the software approach still difficult.

An example of a more refined criterion of EPA measurement can be
derived from an attentive comparison between the Fig.~\ref{fotito1}c
 and \ref{fotito1}d. In spite of the fact that both objects present
 the characteristic belt of bipolar planetary nebulae, in the case
 of NGC\,5189 the structures that appear in both sides of the
 apparent belt do not have hemispherical appearance as the case shown
 in Fig.~\ref{fotito1}d. The orientation  marked in the figure takes
into account that the brilliant knot at the E of the main body's
external part, is of totally nebular appearance when the original
image is inspected. Considering that the criterion for determination
is based mainly on the external morphology, the major axis of the PN
is well determined. This complex object represents a limit case
where the measured EPA could show the largest scatter when employing
other criteria. But as we show in Section 7, our EPA values do not
show significant deviations from the values obtained by CAM98 for
objects present in both samples.

\section{TRANSFORMATION TO THE GALACTIC SYSTEM}

The EPAs are measured relative to the equatorial coordinates system.
In order to transform this position angle to the galactic system we
need to compute the angle $\beta$, which is subtended at the
position of each object by the direction of the equatorial north and
the galactic north.

\begin{displaymath}
\sin \beta = \frac{ \cos \delta_{0} \sin (\alpha_{0} - \alpha) } {\cos b}
\end{displaymath}

Where $\delta_{0}$ and $\alpha_{0}$ are the equatorial coordinates
of the galactic North Pole, $\alpha$ and $b$ are the equatorial
right ascension and galactic latitude coordinates respectively (both
coordinates were extracted from CGPN2000). Then the galactic
position angle (GPA) is defined as the position angle of the major
axis of the apparent nebular elongation, measured from the direction
of the galactic north towards the east (Table~4).

\begin{displaymath}
GPA = EPA - \beta
\end{displaymath}

\begin{figure}
\includegraphics{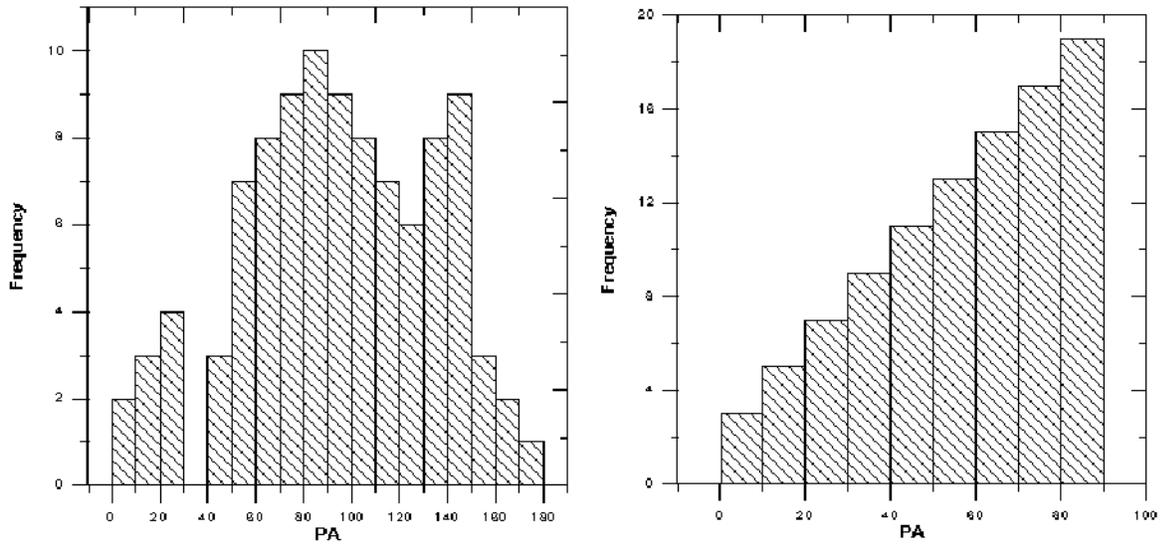}
\vspace{16cm} \caption{Test distribution of position angles,
simulating two peaks not very near to the Galactic plane with PA
computed in the interval $[0^{\circ},180^{\circ}]$ (left), and the
corresponding inferred distribution if the values are reported as
some authors have done, in the interval $[0^{\circ},90^{\circ}]$
(right).} \label{fotito2}
\end{figure}

We used the same convention that Phillips (1997) used for measuring
the EPAs, in the interval $[0^{\circ}, 180^{\circ}]$. The values are
reported and used here in the same range, but some authors (e.g.
CAM98) report them to the interval $[0^{\circ}, 90^{\circ}]$
assuming a priori a symmetric problem, in the sense that a preferred
orientation of the PNe elongations very nearly to the galactic
equator will be easier to detect. The problem with this treatment of
the EPA is that it would blur any other preferred orientation that
is not very near to the galactic equator (Fig.~\ref{fotito2}).

\section{STUDIED REGIONS}

The orientations for individual planetary nebulae is presented in
Table~4, which is also available on request to the authors.  One
difference with previous studies is that, in order to avoid
information degradation by projection effects, we did not include in
our study objects with $\mid b \mid
>20^{\circ}$. The number of objects far from the
galactic plane is not statistically significant, although the
projection effect with respect to the galactic equatorial plane
should be considered in any future larger sample to be analyzed. To
study the orientation of the sample, we selected four angular
sectors: those defined by the galactic center and anti-center, and
the perpendicular directions (Table~1).

The regions were defined in angular range in such a way that there
was enough number of objects to do a statistical analysis in at
least the region towards the galactic center (262 objects). But they
should not be so large in angular extent that any absolute
orientation respect to the galactic system would be deleted by the
superposition of different projections to the observer. It must be
taken into account, that the projection effects at the position of
the observer would mask the effect of some preferred orientations if
the data are compiled and analyzed as a whole set without any regard
to the position respect to the observer and the galactic center. All
the previous works have treated the data as a whole set, and we show
that in this way important information could have been or might be
overlooked in the future.

\section{RESULTS}

\begin{figure}
\includegraphics{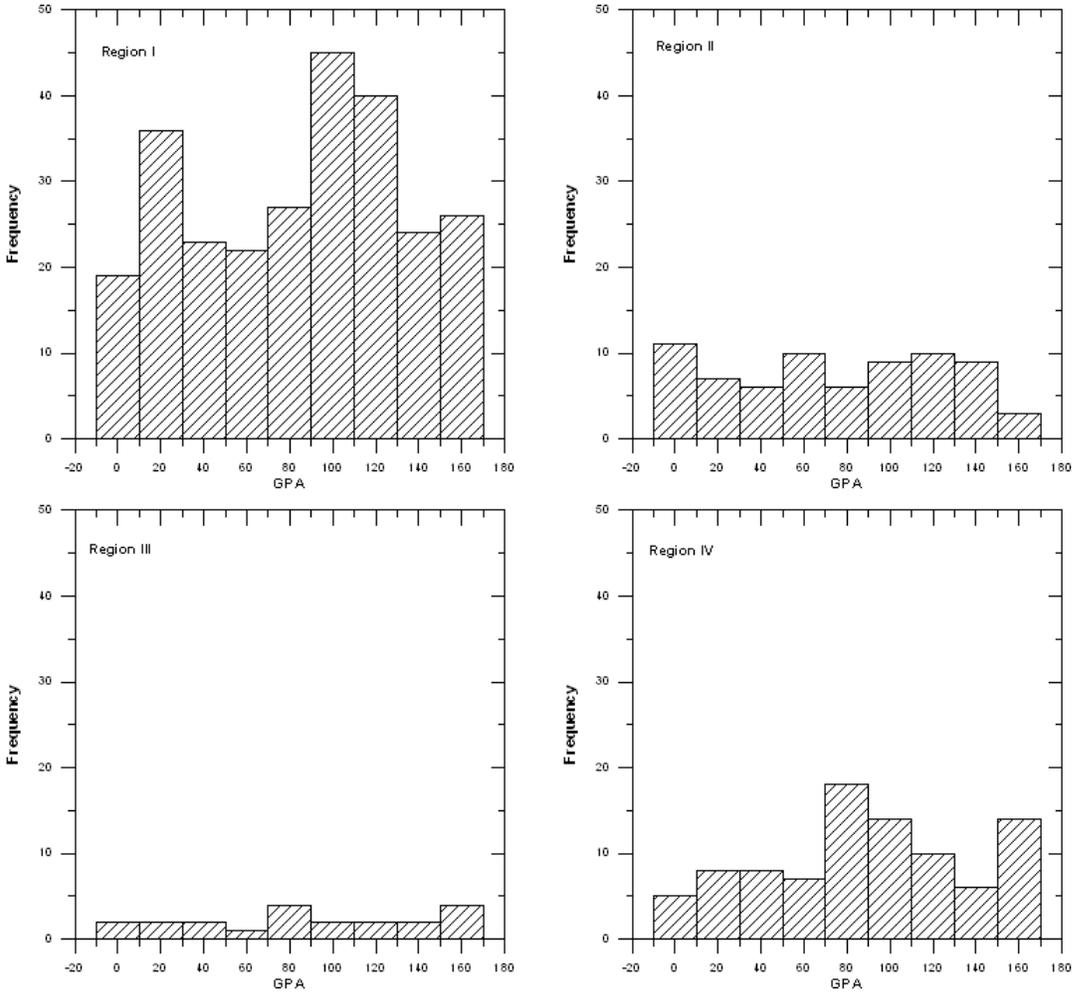}
\vspace{16cm} \caption{Distribution of GPA of the regions presented
in Table~1.} \label{fotito3}
\end{figure}

The region that shows a distribution of GPA with some possible
non-random characteristics is the region with center towards the
galactic center. In the other regions, the low number of objects
precludes detection of any clear trend, although we can rule out any
strong correlation that involves the majority of the PNe in the
sample. The distribution of GPA of the regions are presented in
Fig.~\ref{fotito3}.

If the distribution of the \textbf{N} objects of the sample in each
sector were random, with the same probability of finding an
elongated object with its axis towards any PA, when  \textbf{j}
angular boxes are defined in the range $[0^{\circ}, 180^{\circ}]$,
it could be expected an average population of \textbf{N/j} objects
per bin. The frequency of all the possible values would be an
angular positions distribution, with a dispersion $ \mathbf{\sigma =
(N/j) ^ {1/2}}$. In Table~2 we show the expected average per angular
bin and the dispersion for each region, and the number of peaks
observed above $ 2 \mathbf { \sigma}$ and $ 1 \mathbf{ \sigma}$.

The barycenters calculated for each peak, determined within a window
of  $60 ^{\circ}$, are summarized in Table~2. A thorough test can be
performed by applying a simple binomial test, as proposed by Siegel
(1956) and applied by Hutsem\'ekers (1998). It gives the probability
$P_{s}$ that a random distribution has $ L_{s}$ angles (of a set of
$N$ objects) in the interval $[P_{m} - \alpha, P_{m} + \alpha]$. In
our case $P_{m}$ is the peak barycenter. Such probability is defined
as:

\begin{displaymath}
P_{s}= \sum_{l=l_{s}}^N \left(\frac{\alpha}{90}\right)^{l}
\left(1-\frac{\alpha}{90}\right)^{N-l} {N \choose l}
\end{displaymath}

Table~3 shows the probability of randomness for each observed peak
using $\alpha = 20^{\circ}$ and $\alpha = 10^{\circ}$ (note that
when $\alpha = 20^{\circ}$  the probability that the right peak of
region I is random, is only $P=0.016$). In region I, the
distribution has two peaks with a separation of $\sim 80^{\circ}$
(or $\sim 100^{\circ}$ in the opposite sense) in the directions
defined by the peaks. {\it It is important to remark that the
separate subsamples of PNe from DSS2, CAM98 and MASH each one shows
at the region I, an apparent peak in $GPA \simeq 100^{\circ}$.}

\begin{figure}
\includegraphics{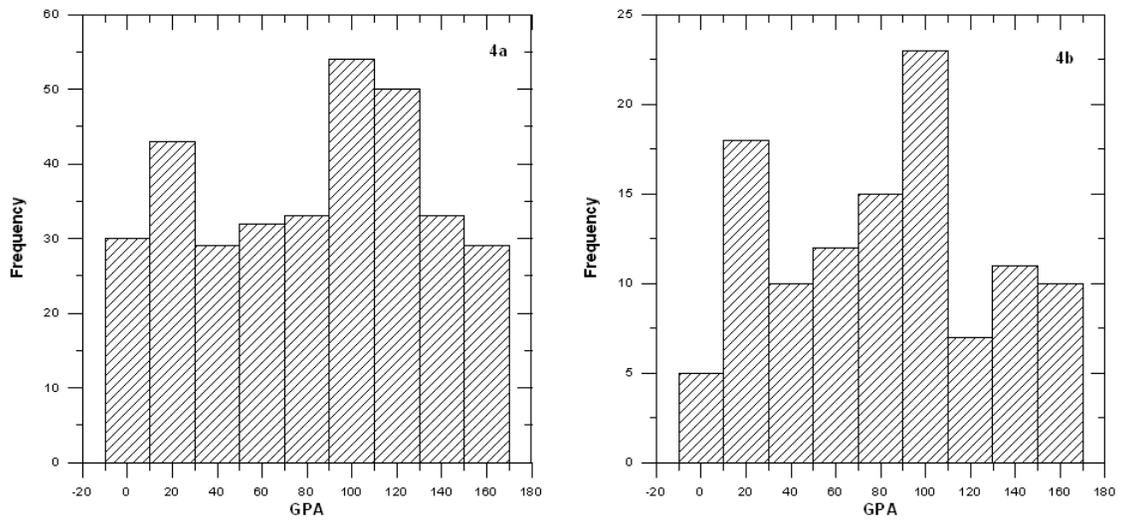} \vspace{16cm} \caption{Added distribution for regions
$I+II$ (a). Inversion of the measurement of regions $III+IV$,
considering the projection effect over the observer (b). Note that
both distributions show a maximum value around 100-110 degrees, and
a possible secondary excess at 20-30 degrees.} \label{fotito4}
\end{figure}

Following the clue provided by some similarity of the results for
regions III and IV, we added the distributions found for regions I
and II (Fig.~\ref{fotito4}a) and for regions III and IV
(Fig.~\ref{fotito4}b). The two distributions in Fig.~\ref{fotito4}b
show right peak over $2\sigma$ and left  peak over $1\sigma$, in
both cases with a peak separation of $80^{\circ}$. Here the
separation is considered as the minimum angular distance between
both peaks. Moreover as both distributions are in opposite sides of
the sky, is natural to measure the separation in opposite sense. It
should be noted that the position angles are being measured with the
NESW convention, without any regard to the absolute orientation of
the objects respect to the observer. Then if some absolute
orientation would be common to most of the objects and considering
that the observer is in the center of the sky band over which the
objects appear projected, the sense of measuring of the position
angles in opposite regions (e.g. I and II) should be inverted by
making \textbf{180 - GPA} in the opposite regions. Coincidentally,
the transformation of the data in regions III+IV
(Fig.~\ref{fotito4}b) gives a distribution resembling that of
regions I+II: a narrow main peak and a possible secondary one
$80^{\circ}$ before.

We also carried out the Kolmogorov-Smirnov (K-S) test to have an
alternative evaluation of the randomness of the observed
distribution. This requires the calculation of     $D_{max}$: the
absolute value of the maximum deviation between the observed
($S_{N}(x)$ with $N_{1}$ points)  and theoretical ($P(x)$ with
$N_{2}$ points) cumulative distribution function (Press et al.
1992). The significance can be written as the following sum:

\begin{displaymath}
Q_{ks}=2\sum_{j=1}^\infty (-1)^{(j-1)} e^{-2j^2 \lambda^2}
\end{displaymath}

Where $\lambda=(\sqrt{N_{e}}+0.12+0.11/\sqrt{N_{e}})D_{max} $ and
$N_{e}=N_{1}N_{2}/(N_{1}+N_{2})$ that is the effective number of
data points of the samples. This test gave us a probability, that
our data derived from a random distribution, of 63\% for region I,
53\% for region II, 93\% for region III and 28\% for region IV.

As an alternative we tried to perform a variation of K-S test: the
Kuiper (K-P) test which is
 more sensible than K-S test in some kind of circular distributions.

To perform this test we have to calculate $V=D_{+}+D_{-}$, which is
the sum of the maximum distance of $S_{N}(x)$ above and below
$P(x)$. A good approximation for the significance is:

\begin{displaymath}
Q_{kp}=2\sum_{j=1}^\infty (4j^2\lambda^2-1) e^{-2j^2 \lambda^2}
\end{displaymath}

Where $\lambda=(\sqrt{N_{e}}+0.155+0.24/\sqrt{N_{e}})V $

The result that we obtained with K-P test for our region I (with the
strongest signal) is $Q_{KP}=35\%$, which seems to be not a
convincing one. These tests could not be fully applicable to our
kind of data: caution should be taken when applying all varieties of
K-S test, because they lack the ability to discriminate some kind of
distributions. For example, we can consider a probability
distribution showing a narrow hole within which the randomness
probability falls to zero. The existence of even one data point
within the hole would rule out such distribution (because of its
cumulative nature, Press et al. 1992), the K-S test would require
many data points in the narrow hole before signaling a discrepancy.

\begin{figure}
\includegraphics{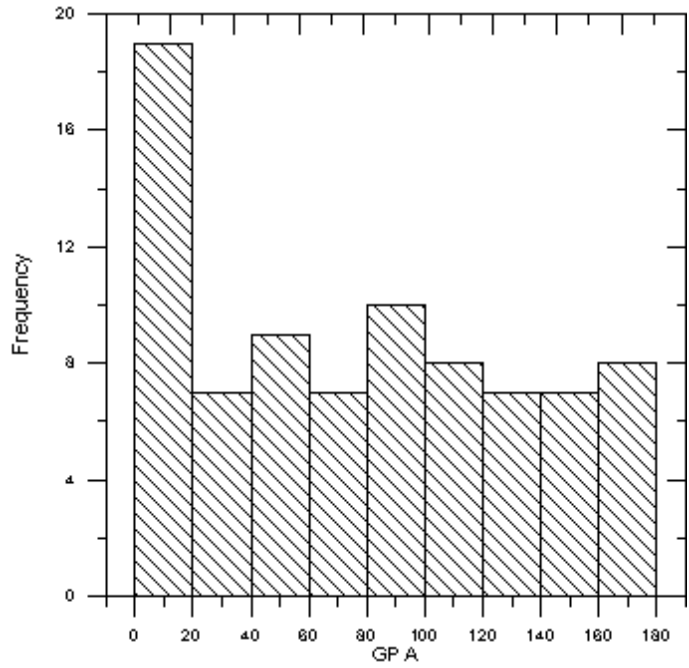}
\vspace{16cm} \caption{ Artificial distribution of GPA, with a clear
signal
 in the first bin.}
\label{fotito5}
\end{figure}

\begin{figure}
\includegraphics{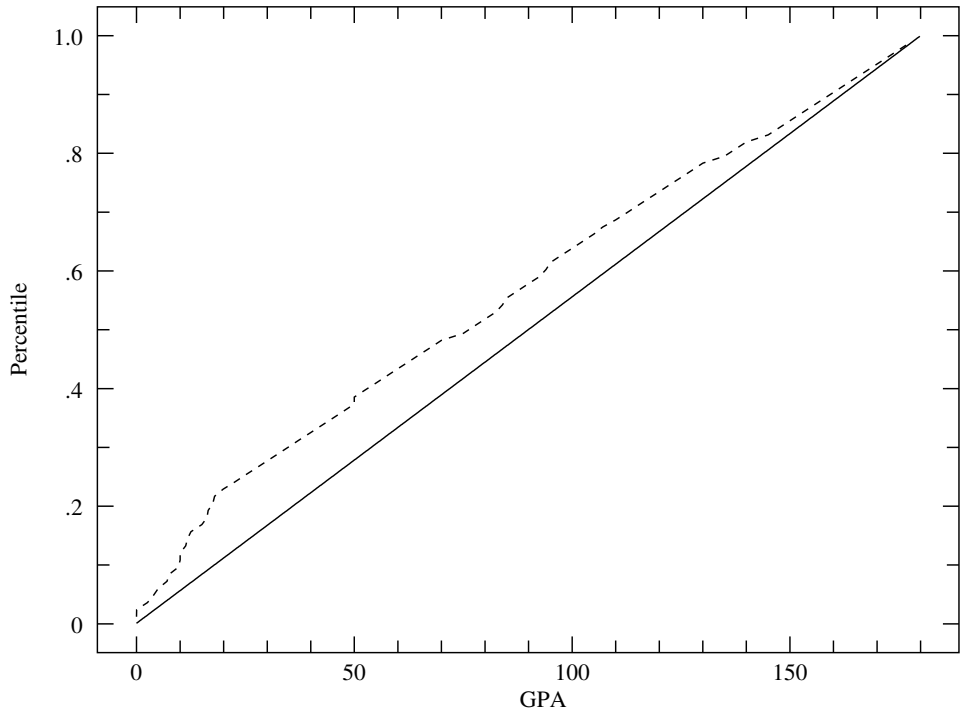}
\vspace{16cm} \caption{ K-P test comparison percentile plot of the
distribution shown in   (Fig.6).
Despite the clear 3 sigma signal presented,
 the largest $D_{max}$ yields, through the K-P test, a
probability of $67\%$.} \label{fotito6}
\end{figure}

To probe this kind of behavior, we generated a random distribution
of 72 GPA and after that we added 10 GPA values distributed in the
first bin (Fig.~\ref{fotito5}). This distribution has an average of
9.1 GPA per bin and a dispersion of $\sigma=3.0$.  In this way, the
first bin shows a signal over $3\sigma$. Now if we apply both tests
to this artificial sample we obtain the next probabilities, that
this data follow a random distribution:
 $Q_{KS} = 17\%$ and $Q_{KP} = 67\%$ (Fig.~\ref{fotito6}).
We have verified that there is no way to generate 67 cases and not
even 17 cases of a $3\sigma$ peak in 100 fully random samples.

Therefore we conclude that although K-S and K-P tests are well
suited to analyze non
 random global trends in samples, they are strongly insensitive to the
 presence of a local non-random feature.

\section{DISCUSSION}

\begin{figure}
\includegraphics{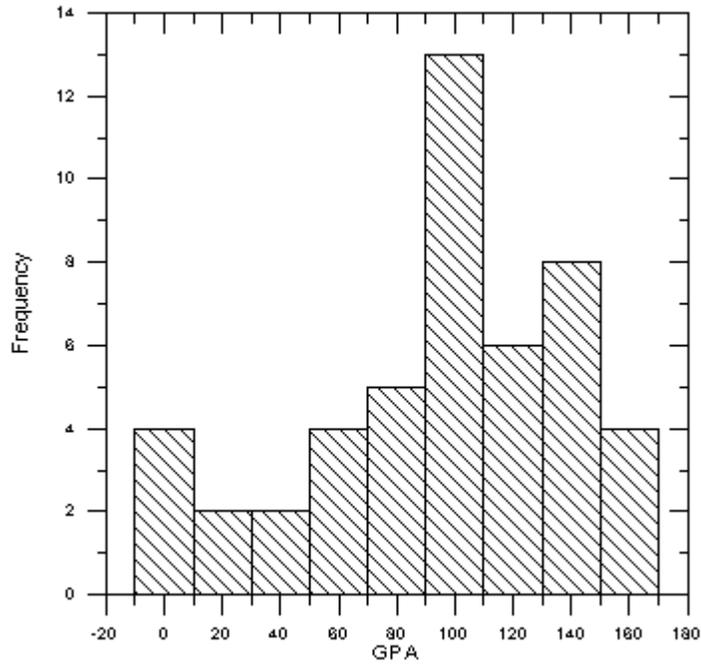}
\vspace{16cm} \caption{ Distribution of GPA of those objects from
the sample of CAM98 that have coordinates in our region I. Note that
it also presents a peak over 2 sigma located at $ GPA \simeq
100^{\circ}$, which was overlooked by the way of analyzing the GPA
distribution in the original work.} \label{fotito7}
\end{figure}

\begin{figure}
\includegraphics{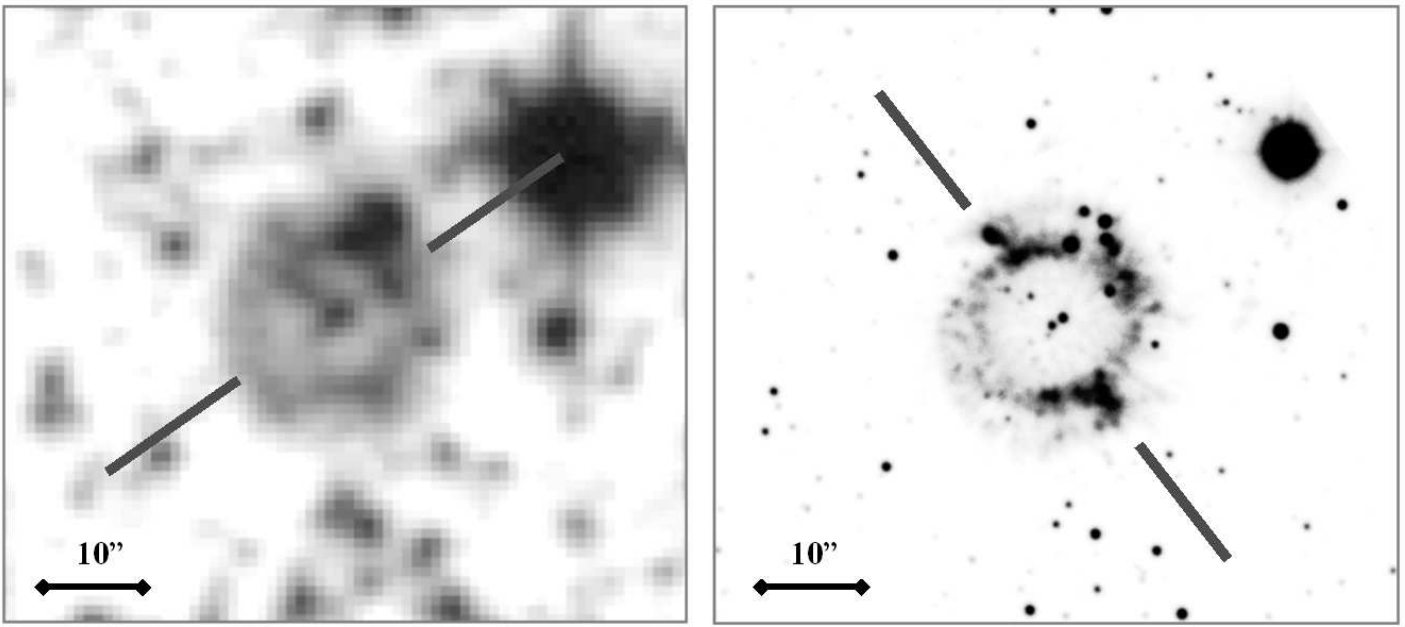}
\vspace{16cm} \caption{ Left: PN TH2-A imaged with DSS at
1.5\arcsec\, resolution, R broad-band.  Right: The same objet imaged
with Gemini-S at 0.7\arcsec\, seeing, through an [OIII]$\lambda$5007
narrow-band filter. The presence of faint blue emission knots can
change the major axis determination from the ellipsoid maximum
diameter to a value 85 degrees apart (this object is not included in
the final sample).
 North is up and East to the left.} \label{fotito8}
\end{figure}

We studied the orientation of all PNe of our sample (without
separating it in regions) and we do not see a clear preferential
orientation of long axis of PN respect to the galactic plane. This
distribution of GPA is similar to that observed by CAM98. We tried
to double check our results by applying our analysis to the data
published by CAM98, performing a suitable analysis to take over the
ambiguity of the position angles (assigned to the range $[0^{\circ},
90^{\circ}]$). More precisely, the sample of CAM98 includes 208
objects (counting the object M2-55 in only one list); 69 of them are
in both samples (the object that CAM98 put in the list of
ellipticals as PC 4, is in fact PB 4); 85 are outside our four
regions; 50 were not measured because they are too small; have
strange morphology or circular appearance; and 4 objects are not
true PN (Bl Cru, CRL 2688, M 1-91, M 1-92). 75$\%$ of the 69 objects
that are in both samples, have their GPA in good agreement
($\pm10^{\circ}$) with our measurement. Moreover, the distribution
of GPA from those objects from CAM98 whose coordinates are in region
I, shows a clear peak over 3 $\sigma$ (Fig.~\ref{fotito7}, 48
objects),
 whose barycenter is in $100^{\circ}\pm15^{\circ}$, a position similar,
 within the uncertainties, to the main peak in our Fig.~\ref{fotito3}.
 This feature could not be detected by CAM98 due to the way they calculated
 the final angles to plot. The differences could be mainly caused by the fact that they
employed deep narrow-band images, as shown in Fig.~\ref{fotito8}.
This figure shows an example of the PN TH2-A DSS image through R
broad-band filter and 1.5\arcsec\, resolution. The EPA was measured
following the ellipsoid apparent major axis. For comparison,
Fig.~\ref{fotito8}-right shows the same object imaged with GMOS at
Gemini-S 8\,m telescope, through a [OIII]$\lambda$5007 filter and
with 0.7\arcsec\, seeing (from D\'{\i}az et al., in preparation).
The presence of faint blue emission knots can change the major axis
determination from the ellipsoid maximum diameter to an
approximately perpendicular angle. Therefore it can be expected that
the measured geometrical properties of the objects change as
distinct ionization layers are imaged through different filters,
more precisely, the left peak (e.g. We 1-4) in our broad band study
could be arose in the measuring of the PN equatorial belt apparent
axis instead of the faint external envelope. This should not occur
in the deep narrow band images of the CAM98 sample and could explain
the absence of the second peak in that sample. Nevertheless, the
agreement between the GPA measurements in both samples is very good
and the presence of the second peak can just be ascribed to the
ability to detect the outer fainter details in the deep narrow band
images, which are usually perpendicular to the bright equatorial
belt structures more easily detected in the DSS imagery.

An interesting test to perform is to check if the distribution of
GPA in the four regions has a distance modulation and consider
objects statistically far and close.  As a first order approach we
separated objects in each region by its angular size: an angular
diameter of 35$\arcsec$ divided the sample in two halves, objects
with large angular size were considered closer than the ones with
small angular size. Assuming an average radius of 0.1\,pc the
separating distance is 1.2\,kpc.  The results that we obtained in
the four regions show that the distribution of GPA for larger PNe
has the same shape than the smaller PNe. So there is not evidence
about a distance effect.

Following the same open minded search we tried to relate the PA of
the long axis of PNe with respect to the Gould Belt, the plane
defined by nearby stars of young population, mainly O and B stars
(Cameron et al. 1994). The result found shows a noisy distribution
of PA. Besides, several studies have related bipolar PNe with binary
progenitors (Bond \& Livio 1990) but the antecedents about the
orientations of the orbits of binary stars do not provide any
comparable result. Notwithstanding it must be remarked that a
thorough analysis should be made, keeping in mind the new geometric
considerations carried out in this work.

We could also consider the case in which only the disk population of
the PNe towards the galactic center would show a preferred
orientation. First, we can test the robustness of the result against
the possibility of significant contamination of the PNe sample
toward the galactic center caused by a random orientation of bulge
PNe. If the bulge PNe contamination were as high as 50\%, then more
than half of the disk PNe would need to have a preferred orientation
near to the galactic plane, to be detected and identified in the EPA
distribution.  Moreover, it would be necessary to disentangle the
bulge objects by their radial velocities in order to verify if the
preferred orientation for the disk PNe is actually much larger than
the one reported here.

Regarding the physical causes, Melnick \& Harwitt (1974) mentioned
the compression of the PNe shells resulting from motion through the
ISM is seldom observed, remarking that the time scale of the PN
expansion ($10^4$\,years) would be large enough to show systematic
off-centering of the progenitor stars, which heretofore has been
detected in a few lopsided objects (Borkowski, Sarazin \& Soker
1990) and could be a dominant factor at the very faintest outer
envelopes where the nebular density falls below a critical limit of
N$_H\sim40$\,cm$^{-3}$. Undoubtedly the most considered hypothesis
has been that the PNe could eventually expand more in the direction
of the ambient or galactic magnetic field field force lines, which
are approximately deployed along the Milky Way plane (e.g. Phillips
1997). The typical energy density of the interstellar magnetic field
is lower than $B^2/8\Pi\approx1.5E-11$\,erg\,cm$^{-3}$, whereas the
energy density (thermal, excitation and ionization) is usually
larger than $E-9$\,erg\,cm$^{-3}$, consequently a correlation with
any observable quantity related with galactic magnetic fields (Diaz
\& Weidmann 2008, in preparation) would imply that this are at least
one or two orders of magnitude higher in some regions towards the
galactic center.

\section{FINAL REMARKS}

The data on PA of planetary nebulae show that global
preferred orientations are not dominant, even considering
 a new approach that takes into account the three-dimensional problem.
 Besides the conclusions hold by the results, we presented here some important
geometrical and analytical considerations that have been overlooked
by the previous works on the subject.

It is worth to mention that there could be a preferred orientation
of some PNe in the zones near to the galactic center, and the
corresponding distribution of galactic position angles could have a
peak not exactly aligned with the Milky Way plane: the PNe towards
the galactic center have an orientation distribution with a possible
narrow peak near to the galactic plane ($GPA \simeq 100^{\circ}$),
with a randomness probability as small as $0.01$. We remark that we
did not exclude any object within the selection criteria and
galactic coordinates in the range $-30< l <30$ and $-20< b <20$, and
no special attention was given to the possibility that some objects
may belong to the galactic bulge population and the consequences
over the detected trend are unknown beyond the qualitative
discussion of the previous section.

Furthermore a larger and deeper galactic center sample, optimally
observed at NIR wavelengths, should be analyzed in order to
thoroughly assess the reality of this preferred orientation. The
possible implications are of broad astrophysical interest and we
hope that they stimulate more detailed studies, in particular,
larger samples of galactic PNe towards the galactic center should be
studied as soon as they become available.

\section*{Acknowledgments}

This work was partially supported by Gustavo Carranza through the
Agencia Cordoba Ciencia and CONICET of Argentina. We acknowledge the
referee for his useful suggestions, in particular the hint to use
the MASH survey. RD acknowledges fruitful discussions about the
original manuscript with Romano Corradi in 2005, and thanks Percy
Gomez for a critical reading of the manuscript. This research has
made use of Aladin and the Multimission Archive at the Space
Telescope Science Institute (MAST). STScI is operated by the
Association of Universities for Research in Astronomy, Inc., under
NASA contract NAS5-26555. Support for MAST for non-HST data is
provided by the NASA Office of Space Science via grant NAG5-7584 and
by other grants and contracts. The Gemini Observatory is operated by
the Association of Universities for Research in Astronomy, Inc.,
under a cooperative agreement with the NSF on behalf of the Gemini
partnership: NSF (USA), PPARC (United Kingdom), NRC (Canada), ARC
(Australia), CONICET (Argentina), CNPq (Brazil) and CONICYT (Chile).

%
%

\clearpage

\begin{table*}
 \centering
 \begin{minipage}{140mm}
  \caption{Sampled galactic sectors.}
  \begin{tabular}{@{}lccc@{}}
Sector & l range [$^{\circ}$] & b range [$^{\circ}$] & Number of
objects\\
Galactic center (Region I)         &  -30 a 30   &   -20 a 20   &   262  \\
Local motion apex (Region II)                 &   60 a 120  &   -20 a 20   &   71  \\
Galactic anti-center (Region III)             &  150 a 210  &   -20 a 20   &   21  \\
Local motion antapex (Region IV)              &  240 a 300  &   -20 a 20   &   90  \\
\end{tabular}
\end{minipage}
\end{table*}


\begin{table*}
 \centering
 \begin{minipage}{140mm}
  \caption{GPA Barycenters of the peaks observed in Fig. 3, in degrees.}
  \begin{tabular}{@{}lcccccc@{}}
Region & Expected Average & Dispersion & $2\sigma$ peak? & Baryc.1 & Baryc.2 & Separation\\
 & (number per bin) & & & (main peak) &
 & \\
I   &    29.1   &   5.4   &   yes   &   20 $\pm$ 16  &   101 $\pm$ 16   &  81  \\
II  &    7.9    &   2.8   &   -   &   6 $\pm$ 17   &   -              &  -  \\
III &    2.3    &   1.5   &   -   &   79 $\pm$ 15  &   158 $\pm$ 16   &  79  \\
IV  &    10.0   &   3.2   &   yes   &   84 $\pm$ 15  &   158 $\pm$ 16   &  74  \\
\end{tabular}
\end{minipage}
\end{table*}


\begin{table*}
 \centering
 \begin{minipage}{140mm}
  \caption{Randomness probabilities of each sector
with a peak width of $20^{\circ}  (\alpha=10^{\circ})$ and
$40^{\circ} (\alpha=20^{\circ})$.}
  \begin{tabular}{@{}lccccc@{}}
Regions & (Left peak)$_{20}$  & (Right peak)$_{20}$ & (Left
peak)$_{10}$ & (Right peak)$_{10}$  \\
I       &  0.841   &   0.003   &     0.077   &  0.0004    \\
II      &  0.215   &  -        &     0.091   &  -         \\
III     &  0.317   &   0.317   &     0.199   &   0.418    \\
IV      &  0.003   &   0.732   &     0.020   &   0.039    \\
\end{tabular}
\end{minipage}
\end{table*}

\clearpage

%

\begin{deluxetable}{lccccc}
\tablecaption{PN Data.}
\tablehead{
\colhead{Name} &
\colhead{l [$^{\circ}$]} &
\colhead{ b [$^{\circ}$]} &
\colhead{EPA [$^{\circ}$]} &
\colhead{GPA[$^{\circ}$]} &
\colhead{Source}  \\
}
\startdata
G000.0-01.8   &   0   &   -1.8   &   163   &   42   &   MASH    \\
M 2-19   &   0.2   &   -1.9   &   110   &   170   &   CAM98    \\
G000.2-03.4   &   0.2   &   -3.4   &   100   &   160   &   MASH    \\
IC 4634   &   0.3   &   12.2   &   153   &   27   &   DSS2    \\
G000.3+07.3   &   0.3   &   7.3   &   90   &   145   &   MASH    \\
G000.3+04.5   &   0.3   &   4.5   &   160   &   36   &   MASH    \\
K 1-4   &   1   &   1.9   &   155   &   33   &   DSS2    \\
G001.2-05.6   &   1.2   &   -5.6   &   138   &   19   &   MASH    \\
He 2-262        &   1.3   &   2.2   &   21   &   78   &   HST    \\
G001.5-02.4   &   1.5   &   -2.4   &   60   &   120   &   MASH    \\
SwSt 1          &   1.6   &   -6.7   &   126   &   8   &   HST    \\
H 1-55          &   1.7   &   -4.5   &   90   &   151   &   HST    \\
G001.8-05.0   &   1.8   &   -5   &   55   &   116   &   MASH    \\
G001.9+02.1   &   1.9   &   2.1   &   80   &   138   &   MASH    \\
IC 4776   &   2   &   -13.4   &   34   &   100   &   CAM98    \\
G002.0+06.6   &   2   &   6.6   &   60   &   116   &   MASH    \\
G002.0+01.5   &   2   &   1.5   &   115   &   173   &   MASH    \\
G002.0-03.2   &   2   &   -3.2   &   95   &   155   &   MASH    \\
G002.1-02.4   &   2.1   &   -2.4   &   130   &   10   &   MASH    \\
G002.1-02.8   &   2.1   &   -2.8   &   15   &   75   &   MASH    \\
H 1-54          &   2.1   &   -4.2   &   137   &   18   &   HST    \\
G002.2+05.8   &   2.2   &   5.8   &   5   &   61   &   MASH    \\
G002.2-01.2   &   2.2   &   -1.2   &   40   &   99   &   MASH    \\
G002.3+01.7   &   2.3   &   1.7   &   35   &   93   &   MASH    \\
Cn 1-5          &   2.3   &   -9.5   &   162   &   45   &   HST    \\
NGC 6369   &   2.4   &   5.9   &   135   &   44   &   DSS2    \\
G002.4+03.5   &   2.4   &   3.5   &   46   &   103   &   MASH    \\
G002.4+01.1   &   2.4   &   1.1   &   62   &   120   &   MASH    \\
G002.4-05.0   &   2.4   &   -5   &   15   &   76   &   MASH    \\
H 2-37          &   2.4   &   -3.4   &   71   &   131   &   HST    \\
G002.5+04.8   &   2.5   &   4.8   &   16   &   73   &   MASH    \\
M 1-42   &   2.7   &   -4.8   &   21   &   83   &   CAM98    \\
G002.8-04.1   &   2.8   &   -4.1   &   65   &   126   &   MASH    \\
Te 1567         &   2.8   &   1.8   &   0   &   58   &   HST    \\
G002.9-03.0   &   2.9   &   -3   &   15   &   75   &   MASH    \\
G003.0-01.7   &   3   &   -1.7   &   140   &   20   &   MASH    \\
G003.1+05.2   &   3.1   &   5.2   &   135   &   12   &   MASH    \\
G003.1-01.6   &   3.1   &   -1.6   &   100   &   160   &   MASH    \\
Hb  4           &   3.2   &   2.9   &   140   &   18   &   HST    \\
G003.3-01.6   &   3.3   &   -1.6   &   30   &   90   &   MASH    \\
IC 4673   &   3.5   &   -2.4   &   126   &   7   &   DSS2    \\
G003.5+04.5   &   3.5   &   4.5   &   35   &   92   &   MASH    \\
G003.5+02.6   &   3.5   &   2.6   &   45   &   103   &   MASH    \\
G003.6-03.0   &   3.6   &   -3   &   145   &   25   &   MASH    \\
H 2-15          &   3.8   &   5.3   &   R   &   ...   &   HST    \\
H 1-59          &   3.9   &   -4.4   &   71   &   132   &   HST    \\
G004.0-02.6   &   4   &   -2.6   &   115   &   175   &   MASH    \\
G004.0-02.7   &   4   &   -2.7   &   92   &   152   &   MASH    \\
G004.1-03.3   &   4.1   &   -3.3   &   85   &   145   &   MASH    \\
G004.2-02.5   &   4.2   &   -2.5   &   140   &   20   &   MASH    \\
G004.5+06.0   &   4.5   &   6   &   60   &   117   &   MASH    \\
H 2-12          &   4.5   &   6.8   &   ...   &   ...   &   HST    \\
G004.8-01.1   &   4.8   &   -1.1   &   32   &   92   &   MASH    \\
H 2-25          &   4.9   &   2.1   &   41   &   99   &   HST    \\
M 1-25          &   4.9   &   4.9   &   40   &   97   &   HST    \\
G005.0+02.2   &   5   &   2.2   &   166   &   44   &   MASH    \\
G005.4-03.4   &   5.4   &   -3.4   &   168   &   49   &   MASH    \\
G005.9-09.8   &   5.9   &   -9.8   &   140   &   24   &   MASH    \\
M 1-28   &   6   &   3.1   &   14   &   73   &   DSS2    \\
G006.1+03.8   &   6.1   &   3.8   &   0   &   58   &   MASH    \\
G006.1+01.5   &   6.1   &   1.5   &   65   &   124   &   MASH    \\
M 1-20          &   6.2   &   8.4   &   102   &   159   &   HST    \\
G006.3+01.7   &   6.3   &   1.7   &   170   &   49   &   MASH    \\
G006.4-03.4   &   6.4   &   -3.4   &   147   &   28   &   MASH    \\
G006.4-05.5   &   6.4   &   -5.5   &   140   &   22   &   MASH    \\
G006.5+08.7   &   6.5   &   8.7   &   54   &   111   &   MASH    \\
G006.5-03.9   &   6.5   &   -3.9   &   20   &   81   &   MASH    \\
M 3-15          &   6.8   &   4.2   &   122   &   0   &   HST    \\
G007.1+07.3   &   7.1   &   7.3   &   136   &   13   &   MASH    \\
G007.1+04.9   &   7.1   &   4.9   &   118   &   176   &   MASH    \\
G007.1-05.0   &   7.1   &   -5   &   65   &   127   &   MASH    \\
G007.3+01.7   &   7.3   &   1.7   &   30   &   89   &   MASH    \\
G007.4+01.7   &   7.4   &   1.7   &   13   &   72   &   MASH    \\
M 2-34   &   7.8   &   -3.7   &   177   &   59   &   DSS2    \\
G007.8+04.3   &   7.8   &   4.3   &   80   &   138   &   MASH    \\
H 1-65          &   7.9   &   -4.4   &   R   &   ...   &   HST    \\
NGC 6445   &   8   &   3.9   &   149   &   110   &   DSS2    \\
M 1-40   &   8.3   &   -1.1   &   44   &   105   &   CAM98    \\
G008.3+09.6   &   8.3   &   9.6   &   120   &   177   &   MASH    \\
He 2-260        &   8.3   &   6.9   &   85   &   143   &   HST    \\
G008.4-02.8   &   8.4   &   -2.8   &   85   &   146   &   MASH    \\
G008.7-04.2   &   8.7   &   -4.2   &   135   &   17   &   MASH    \\
G009.0-02.2   &   9   &   -2.2   &   15   &   76   &   MASH    \\
G009.0-02.4   &   9   &   -2.4   &   18   &   79   &   MASH    \\
G009.4-01.2   &   9.4   &   -1.2   &   53   &   114   &   MASH    \\
NGC 6309   &   9.6   &   14.8   &   56   &   114   &   DSS2    \\
A 41   &   9.6   &   10.5   &   143   &   21   &   DSS2    \\
G009.8-01.1   &   9.8   &   -1.1   &   10   &   71   &   MASH    \\
G009.9+04.5   &   9.9   &   4.5   &   60   &   119   &   MASH    \\
G010.0-01.5   &   10   &   -1.5   &   65   &   126   &   MASH    \\
NGC 6537   &   10.1   &   0.7   &   38   &   99   &   DSS2    \\
G010.2+00.3   &   10.2   &   0.3   &   50   &   110   &   MASH    \\
M 2-9   &   10.8   &   18.1   &   179   &   57   &   DSS2    \\
IC 4732         &   10.8   &   -6.5   &   R   &   ...   &   HST    \\
NGC 6578        &   10.8   &   -1.8   &   144   &   25   &   HST    \\
G011.0+01.4   &   11   &   1.4   &   60   &   120   &   MASH    \\
M 2-13   &   11.1   &   11.5   &   40   &   98   &   CAM98    \\
DeHt 10   &   11.4   &   17.9   &   175   &   53   &   DSS2    \\
NGC 6567        &   11.8   &   -0.7   &   107   &   168   &   HST    \\
G011.9+07.3   &   11.9   &   7.3   &   165   &   44   &   MASH    \\
G012.1+02.8   &   12.1   &   2.8   &   150   &   30   &   MASH    \\
PM 1-188        &   12.2   &   4.9   &   R   &   ...   &   HST    \\
G012.5+04.3   &   12.5   &   4.3   &   113   &   173   &   MASH    \\
G013.1+05.0   &   13.1   &   5   &   85   &   145   &   MASH    \\
M 1-33          &   13.1   &   4.2   &   30   &   90   &   HST    \\
G013.6-04.6   &   13.6   &   -4.6   &   90   &   152   &   MASH    \\
We 4-5   &   13.7   &   -15.3   &   141   &   27   &   DSS2    \\
SaWe 3   &   13.8   &   -2.8   &   133   &   15   &   DSS2    \\
G014.6+02.3   &   14.6   &   2.3   &   50   &   110   &   MASH    \\
G014.6+01.0   &   14.6   &   1   &   175   &   56   &   MASH    \\
A 44   &   15.6   &   -3   &   131   &   13   &   DSS2    \\
M 1-39          &   15.9   &   3.4   &   80   &   140   &   HST    \\
M 1-54   &   16   &   -4.3   &   104   &   167   &   DSS2    \\
G016.0-07.6   &   16   &   -7.6   &   135   &   18   &   MASH    \\
G016.3-02.3   &   16.3   &   -2.3   &   168   &   50   &   MASH    \\
G016.4-00.9   &   16.4   &   -0.9   &   139   &   20   &   MASH    \\
G016.6+03.1   &   16.6   &   3.1   &   90   &   151   &   MASH    \\
G018.0-02.2   &   18   &   -2.2   &   56   &   118   &   MASH    \\
G018.5-01.6   &   18.5   &   -1.6   &   160   &   42   &   MASH    \\
DeHt 3   &   19.4   &   -13.6   &   13   &   73   &   DSS2    \\
CTS 1   &   19.8   &   5.6   &   102   &   163   &   CAM98    \\
G020.4+02.2   &   20.4   &   2.2   &   115   &   176   &   MASH    \\
M 1-51   &   20.9   &   -1.1   &   29   &   92   &   CAM98    \\
M 3-55   &   21.7   &   -0.6   &   57   &   120   &   CAM98    \\
M 3-28   &   21.8   &   -0.5   &   2   &   64   &   DSS2    \\
M 1-57   &   22.1   &   -2.4   &   137   &   20   &   CAM98    \\
M 1-58          &   22.1   &   -3.2   &   78   &   140   &   HST    \\
MaC 1-13   &   22.5   &   1   &   26   &   88   &   DSS2    \\
G023.4+00.7   &   23.4   &   0.7   &   145   &   27   &   MASH    \\
M 1-59   &   23.9   &   -2.3   &   116   &   179   &   CAM98    \\
M 2-40   &   24.1   &   3.8   &   80   &   142   &   CAM98    \\
M 4-9   &   24.2   &   5.9   &   167   &   50   &   DSS2    \\
Pe 1-17   &   24.3   &   -3.3   &   45   &   108   &   CAM98    \\
G024.4-03.5   &   24.4   &   -3.5   &   60   &   122   &   MASH    \\
A 60   &   25   &   -11.7   &   103   &   166   &   DSS2    \\
IC 1295   &   25.4   &   -4.7   &   69   &   132   &   CAM98    \\
NGC 6818   &   25.8   &   -17.9   &   13   &   79   &   CAM98    \\
Pe 1-14   &   25.9   &   -0.9   &   45   &   108   &   CAM98    \\
G026.2-03.4   &   26.2   &   -3.4   &   7   &   69   &   MASH    \\
G026.4+02.7   &   26.4   &   2.7   &   25   &   87   &   MASH    \\
G026.9-00.7   &   26.9   &   -0.7   &   56   &   118   &   MASH    \\
A 49   &   27.3   &   -3.4   &   31   &   94   &   CAM98    \\
DeHt 2   &   27.6   &   16.9   &   44   &   107   &   DSS2    \\
G027.6-00.8   &   27.6   &   -0.8   &   40   &   102   &   MASH    \\
G027.8+02.7   &   27.8   &   2.7   &   50   &   112   &   MASH    \\
WeSb 3   &   28   &   10.3   &   145   &   31   &   DSS2    \\
Pe 1-20   &   28.2   &   -4   &   12   &   75   &   CAM98    \\
K 3-2          &   28.6   &   5.2   &   ...   &   ...   &   HST    \\
G028.7-03.2   &   28.7   &   -3.2   &   40   &   102   &   MASH    \\
A 48   &   29   &   0.5   &   37   &   99   &   DSS2    \\
NGC 6751   &   29.2   &   -5.9   &   81   &   144   &   DSS2    \\
G029.8+00.5   &   29.8   &   0.5   &   20   &   82   &   MASH    \\
A 68   &   60   &   -4.3   &   13   &   72   &   CAM98    \\
NGC 6886   &   60.1   &   -7.7   &   54   &   111   &   CAM98    \\
K 3-45   &   60.5   &   -0.3   &   18   &   78   &   CAM98    \\
NGC 6853   &   60.8   &   -3.7   &   129   &   6   &   DSS2    \\
He 2-437   &   61.3   &   3.6   &   77   &   138   &   DSS2    \\
M 1-91   &   61.4   &   3.6   &   77   &   138   &   CAM98    \\
NGC 6905   &   61.4   &   -9.6   &   161   &   36   &   DSS2    \\
M 2-48   &   62.4   &   -0.3   &   67   &   125   &   DSS2    \\
NGC 6720   &   63.1   &   14   &   57   &   123   &   DSS2    \\
M 1-92   &   64.1   &   4.3   &   130   &   11   &   CAM98    \\
BD+30 3639      &   64.8   &   5   &   R   &   ...   &   HST    \\
We 1-9   &   65.1   &   -3.5   &   70   &   126   &   DSS2    \\
He 1-6   &   65.2   &   -5.7   &   122   &   177   &   DSS2    \\
He 2-459        &   68.4   &   -2.7   &   57   &   113   &   HST    \\
M 1-75   &   68.8   &   0   &   152   &   28   &   DSS2    \\
K 3-46   &   69.2   &   3.8   &   110   &   168   &   DSS2    \\
NGC 6894   &   69.4   &   -2.6   &   145   &   20   &   DSS2    \\
K 3-58   &   69.6   &   -3.9   &   83   &   137   &   DSS2    \\
M 3-35   &   71.6   &   -2.3   &   48   &   103   &   CAM98    \\
K 3-57   &   72.1   &   0.1   &   33   &   90   &   CAM98    \\
A 74   &   72.7   &   -17.1   &   52   &   99   &   DSS2    \\
K 3-76   &   73   &   -2.4   &   134   &   9   &   CAM98    \\
GM 1-11   &   73   &   -2.2   &   65   &   119   &   DSS2    \\
NGC 6881   &   74.5   &   2.1   &   139   &   16   &   CAM98    \\
Anon   &   75.6   &   4.3   &   109   &   166   &   DSS2    \\
A 69   &   76.3   &   1.2   &   54   &   108   &   DSS2    \\
Dd 1   &   78.6   &   5.2   &   81   &   137   &   DSS2    \\
M 4-17   &   79.6   &   5.8   &   118   &   174   &   DSS2    \\
CRL 2688   &   80.1   &   -6.5   &   14   &   63   &   CAM98    \\
A 78   &   81.2   &   -14.9   &   146   &   9   &   DSS2    \\
NGC 6884        &   82.1   &   7.1   &   174   &   51   &   HST    \\
K 4-55   &   84.2   &   1   &   85   &   135   &   DSS2    \\
A 71   &   84.9   &   4.4   &   15   &   68   &   DSS2    \\
Hu 1-2   &   86.5   &   -8.8   &   129   &   172   &   CAM98    \\
We 2-245   &   87.4   &   -3.8   &   137   &   1   &   DSS2    \\
NGC 7048   &   88.7   &   -1.6   &   15   &   60   &   DSS2    \\
NGC 7026   &   89   &   0.3   &   4   &   51   &   DSS2    \\
Sh 1-89   &   89.8   &   -0.6   &   50   &   95   &   DSS2    \\
K 3-84   &   91.6   &   -4.8   &   3   &   45   &   CAM98    \\
We 1-11   &   91.6   &   1.8   &   49   &   95   &   DSS2    \\
K 3-79   &   92.1   &   5.8   &   132   &   1   &   DSS2    \\
M 1-79   &   93.3   &   -2.4   &   88   &   129   &   DSS2    \\
NGC 7008   &   93.4   &   5.4   &   38   &   86   &   DSS2    \\
K 3-83   &   94.5   &   -0.8   &   100   &   142   &   CAM98    \\
A 73   &   95.2   &   7.8   &   38   &   87   &   DSS2    \\
K 3-61   &   96.3   &   2.3   &   164   &   27   &   CAM98    \\
M 2-50   &   97.6   &   -2.4   &   52   &   90   &   CAM98    \\
Me 2-2          &   100   &   -8.8   &   R   &   ...   &   HST    \\
IC 5217   &   100.6   &   -5.4   &   90   &   122   &   CAM98    \\
A 75   &   101.8   &   8.7   &   94   &   137   &   DSS2    \\
A 80   &   102.8   &   -5   &   19   &   48   &   DSS2    \\
A 79   &   102.9   &   -2.3   &   20   &   51   &   DSS2    \\
M 2-51   &   103.2   &   0.6   &   151   &   4   &   DSS2    \\
M 2-52   &   103.7   &   0.4   &   116   &   148   &   DSS2    \\
KLW 8   &   104.1   &   -1.4   &   40   &   70   &   DSS2    \\
NGC 7139   &   104.1   &   7.9   &   47   &   86   &   DSS2    \\
M 2-53   &   104.4   &   -1.6   &   95   &   124   &   DSS2    \\
NGC 7354   &   107.8   &   2.3   &   17   &   45   &   DSS2    \\
IRAS22568+6141   &   110.1   &   1.9   &   121   &   146   &   CAM98    \\
K 1-20   &   110.6   &   -12.9   &   154   &   169   &   DSS2    \\
KjPn 6   &   111.2   &   7   &   4   &   32   &   CAM98    \\
KjPn 8   &   112.5   &   -0.1   &   72   &   91   &   CAM98    \\
k 3-88   &   112.5   &   3.7   &   45   &   66   &   DSS2    \\
A 84   &   112.9   &   -10.2   &   172   &   6   &   DSS2    \\
A 83   &   113.6   &   -6.9   &   34   &   47   &   DSS2    \\
A 82   &   114   &   -4.6   &   46   &   59   &   DSS2    \\
We 2-262   &   116   &   0.1   &   172   &   4   &   DSS2    \\
M 2-55   &   116.2   &   8.5   &   38   &   55   &   DSS2    \\
M 2-56   &   118   &   8.4   &   87   &   100   &   CAM98    \\
A  86   &   118.7   &   8.2   &   115   &   125   &   DSS2    \\
BV 5-1   &   119.3   &   0.3   &   53   &   59   &   DSS2    \\
Hu 1-1   &   119.6   &   -6.1   &   9   &   14   &   CAM98    \\
NGC 40   &   120   &   9.8   &   15   &   24   &   CAM98    \\
K 3-64   &   151.4   &   0.5   &   120   &   77   &   CAM98    \\
IC 351   &   159   &   -15.1   &   3   &   143   &   CAM98    \\
IC 2149   &   166.1   &   10.4   &   73   &   11   &   DSS2    \\
CRL 618   &   166.4   &   -6.5   &   99   &   49   &   DSS2    \\
Pu 2   &   173.5   &   3.2   &   47   &   169   &   DSS2    \\
H 3-29   &   174.2   &   -14.6   &   158   &   107   &   DSS2    \\
Pu 1   &   181.5   &   0.9   &   50   &   170   &   DSS2    \\
WeSb 2   &   183.8   &   5.5   &   33   &   152   &   DSS2    \\
NGC 2371   &   189.1   &   19.8   &   129   &   61   &   DSS2    \\
M 1-7   &   189.8   &   7.8   &   153   &   90   &   DSS2    \\
J 320   &   190.3   &   -17.8   &   174   &   117   &   DSS2    \\
HaWe 8   &   192.5   &   7.2   &   134   &   71   &   DSS2    \\
J 900   &   194.2   &   2.6   &   83   &   21   &   DSS2    \\
NCG 2022   &   196.6   &   -11   &   32   &   152   &   DSS2    \\
A 14   &   197.8   &   -3.4   &   175   &   114   &   DSS2    \\
A 12   &   198.6   &   -6.3   &   15   &   134   &   DSS2    \\
A 19   &   200.7   &   8.4   &   65   &   2   &   DSS2    \\
We 1-4   &   201.9   &   -4.7   &   104   &   42   &   DSS2    \\
A 13   &   204   &   -8.5   &   47   &   165   &   DSS2    \\
K 3-72   &   204.8   &   -3.6   &   139   &   77   &   DSS2    \\
A 21   &   205.1   &   14.2   &   136   &   72   &   DSS2    \\
M 3-2   &   240.3   &   -7.6   &   42   &   160   &   DSS2    \\
M 3-4   &   241   &   2.3   &   141   &   83   &   DSS2    \\
M 3-1   &   242.6   &   -11.6   &   143   &   78   &   DSS2    \\
G242.6-04.4   &   242.6   &   -4.4   &   98   &   37   &   MASH    \\
NGC 2452   &   243.3   &   -1.1   &   116   &   57   &   DSS2    \\
A 29   &   244.5   &   12.5   &   133   &   80   &   DSS2    \\
G244.6-00.3   &   244.6   &   -0.3   &   0   &   121   &   MASH    \\
G249.8+07.1   &   249.8   &   7.1   &   55   &   2   &   MASH    \\
A 26   &   250.3   &   0.1   &   141   &   85   &   DSS2    \\
G250.3-05.4   &   250.3   &   -5.4   &   148   &   88   &   MASH    \\
G250.4-01.3   &   250.4   &   -1.3   &   105   &   48   &   MASH    \\
K 1-21   &   251.1   &   -1.6   &   160   &   103   &   DSS2    \\
K 1-1   &   252.6   &   4.4   &   36   &   163   &   DSS2    \\
K 1-2   &   253.5   &   10.7   &   101   &   52   &   DSS2    \\
VBRC 1   &   257.5   &   0.6   &   118   &   65   &   DSS2    \\
He 2-9        &   258.1   &   -0.4   &   31   &   157   &   HST    \\
He 2-11   &   259.1   &   0.9   &   137   &   86   &   DSS2    \\
He 2-15   &   261.6   &   3   &   33   &   164   &   DSS2    \\
NGC 2818   &   261.9   &   8.5   &   90   &   45   &   DSS2    \\
He 2-7   &   264.1   &   -8.1   &   161   &   105   &   DSS2    \\
Wray 17-20   &   264.4   &   -3.6   &   22   &   150   &   DSS2    \\
NGC 2792   &   265.7   &   4.1   &   146   &   100   &   DSS2    \\
G266.8-04.2   &   266.8   &   -4.2   &   125   &   73   &   MASH    \\
G267.5+04.6   &   267.5   &   4.6   &   142   &   97   &   MASH    \\
G268.6+05.0   &   268.6   &   5   &   50   &   6   &   MASH    \\
G269.3-00.8   &   269.3   &   -0.8   &   80   &   32   &   MASH    \\
NGC 3132   &   272.1   &   12.3   &   167   &   131   &   DSS2    \\
He 2-18   &   273.2   &   -3.7   &   75   &   29   &   DSS2    \\
G273.2-00.3   &   273.2   &   -0.3   &   123   &   79   &   MASH    \\
Lo 4   &   274.3   &   9.1   &   46   &   11   &   DSS2    \\
He 2-37   &   274.6   &   3.5   &   119   &   80   &   DSS2    \\
G274.8-05.7   &   274.8   &   -5.7   &   57   &   9   &   MASH    \\
PB 4   &   275   &   -4.1   &   35   &   170   &   DSS2    \\
G275.0+01.5   &   275   &   1.5   &   13   &   152   &   MASH    \\
G275.1-03.5   &   275.1   &   -3.5   &   137   &   92   &   MASH    \\
He 2-25   &   275.2   &   -3.7   &   14   &   148   &   CAM98    \\
G275.6-00.5   &   275.6   &   -0.5   &   128   &   86   &   MASH    \\
He 2-29   &   275.8   &   -2.9   &   55   &   12   &   DSS2    \\
G276.1-03.3   &   276.1   &   -3.3   &   130   &   86   &   MASH    \\
NGC 2899   &   277.1   &   -3.8   &   117   &   73   &   CAM98    \\
G277.4-00.7   &   277.4   &   -0.7   &   138   &   98   &   MASH    \\
Wray 17-31   &   277.7   &   -3.5   &   90   &   48   &   DSS2    \\
G278.3-04.3   &   278.3   &   -4.3   &   30   &   167   &   MASH    \\
He 2-32   &   278.5   &   -4.5   &   159   &   117   &   DSS2    \\
VBRC 3   &   279   &   -3.2   &   56   &   16   &   DSS2    \\
G279.1-03.1   &   279.1   &   -3.1   &   57   &   16   &   MASH    \\
He 2-36   &   279.6   &   -3.1   &   144   &   105   &   DSS2    \\
G279.6+02.2   &   279.6   &   2.2   &   141   &   105   &   MASH    \\
G281.3-07.1   &   281.3   &   -7.1   &   143   &   100   &   MASH    \\
G282.5-02.1   &   282.5   &   -2.1   &   147   &   111   &   MASH    \\
He 2-48   &   282.9   &   3.8   &   129   &   99   &   DSS2    \\
Hf 4   &   283.9   &   -1.8   &   13   &   160   &   DSS2    \\
ESO 215-04   &   283.9   &   9.7   &   143   &   118   &   DSS2    \\
G284.5+03.8   &   284.5   &   3.8   &   127   &   98   &   MASH    \\
G284.6-04.5   &   284.6   &   -4.5   &   10   &   154   &   MASH    \\
IC 2553   &   285.4   &   -5.3   &   7   &   151   &   CAM98    \\
He 2-52   &   285.5   &   1.5   &   105   &   76   &   CAM98    \\
G285.5-03.3   &   285.5   &   -3.3   &   83   &   50   &   MASH    \\
IC 2448   &   285.7   &   -14.9   &   137   &   88   &   DSS2    \\
He 2-47        &   285.7   &   -2.7   &   ...   &   ...   &   HST    \\
Wray 17-40   &   286.2   &   -6.9   &   102   &   67   &   DSS2    \\
G286.3-03.1   &   286.3   &   -3.1   &   145   &   113   &   MASH    \\
G286.3-00.7   &   286.3   &   -0.7   &   130   &   100   &   MASH    \\
Hf 38   &   288.4   &   0.3   &   94   &   69   &   DSS2    \\
G288.4-01.8   &   288.4   &   -1.8   &   107   &   80   &   MASH    \\
Hf 39   &   288.9   &   -0.8   &   43   &   18   &   DSS2    \\
He 2-57   &   289.6   &   -1.6   &   119   &   95   &   DSS2    \\
Hf 48   &   290.1   &   -0.4   &   136   &   113   &   DSS2    \\
G291.3+03.7   &   291.3   &   3.7   &   55   &   36   &   MASH    \\
G291.3+08.4   &   291.3   &   8.4   &   175   &   158   &   MASH    \\
He 2-64   &   291.7   &   3.7   &   81   &   62   &   CAM98    \\
G291.9-04.0   &   291.9   &   -4   &   44   &   21   &   MASH    \\
G292.4+00.8   &   292.4   &   0.8   &   153   &   134   &   MASH    \\
Wray 16-93   &   292.7   &   1.9   &   134   &   117   &   DSS2    \\
He 2-67   &   292.8   &   1.1   &   120   &   102   &   CAM98    \\
G292.8+00.6   &   292.8   &   0.6   &   168   &   150   &   MASH    \\
He 2-70   &   293.6   &   1.2   &   137   &   121   &   DSS2    \\
Lo 6   &   294.1   &   14.4   &   150   &   140   &   DSS2    \\
NGC 3918   &   294.6   &   4.7   &   138   &   126   &   DSS2    \\
He 2-72   &   294.9   &   -0.6   &   101   &   87   &   DSS2    \\
He 2-71        &   296.5   &   -6.9   &   R   &   ...   &   HST    \\
NGC 3195   &   296.6   &   -20   &   13   &   155   &   CAM98    \\
G297.6-01.6   &   297.6   &   -1.6   &   90   &   79   &   MASH    \\
He 2-76   &   298.2   &   -1.7   &   83   &   75   &   DSS2    \\
NGC 4071   &   298.3   &   -4.8   &   43   &   34   &   DSS2    \\
G298.6-01.5   &   298.6   &   -1.5   &   50   &   41   &   MASH    \\
G298.7-07.5   &   298.7   &   -7.5   &   158   &   147   &   MASH    \\
K 1-23   &   299   &   18.4   &   3   &   179   &   DSS2    \\
G299.0+03.5   &   299   &   3.5   &   140   &   133   &   MASH    \\
HaTr 1   &   299.4   &   -4.1   &   131   &   124   &   DSS2    \\
He 2-82   &   299.5   &   2.4   &   162   &   157   &   DSS2    \\
Bl Cru   &   299.7   &   0.1   &   31   &   25   &   CAM98    \\
Lo 9   &   330.2   &   5.9   &   82   &   120   &   DSS2    \\
He 2-153   &   330.6   &   -2.1   &   41   &   86   &   DSS2    \\
He 2-159   &   330.6   &   -3.6   &   173   &   39   &   DSS2    \\
G330.7-02.0   &   330.7   &   -2   &   95   &   139   &   MASH    \\
Wray 16-189   &   330.9   &   4.3   &   72   &   112   &   DSS2    \\
PC 11           &   331.1   &   -5.8   &   R   &   ...   &   HST    \\
G331.3+01.6   &   331.3   &   1.6   &   145   &   6   &   MASH    \\
He 2-145   &   331.4   &   0.5   &   113   &   156   &   DSS2    \\
He 2-165   &   331.5   &   -3.9   &   36   &   83   &   DSS2    \\
He 2-161   &   331.5   &   -2.7   &   48   &   94   &   DSS2    \\
Mz 3   &   331.7   &   -1   &   10   &   55   &   DSS2    \\
He 2-164   &   332   &   -3.3   &   82   &   129   &   DSS2    \\
G332.0-04.3   &   332   &   -4.3   &   35   &   82   &   MASH    \\
G332.5-02.2   &   332.5   &   -2.2   &   117   &   163   &   MASH    \\
HaTr 6   &   332.8   &   -16.4   &   55   &   119   &   DSS2    \\
He 2-152   &   333.4   &   1.1   &   150   &   14   &   DSS2    \\
HaTr 3   &   333.4   &   -4   &   157   &   26   &   DSS2    \\
MeWe 1-6   &   334.3   &   -1.4   &   125   &   172   &   DSS2    \\
IC 4642   &   334.3   &   -9.3   &   143   &   18   &   DSS2    \\
G334.3-13.4   &   334.3   &   -13.4   &   43   &   102   &   MASH    \\
HaTr 4   &   335.2   &   -3.6   &   94   &   144   &   DSS2    \\
He 2-169   &   335.4   &   -1.1   &   3   &   51   &   DSS2    \\
ESO 330-02   &   335.4   &   9.2   &   4   &   45   &   DSS2    \\
DS 2   &   335.5   &   12.4   &   115   &   154   &   DSS2    \\
Pc14   &   336.2   &   -6.9   &   106   &   159   &   CAM98    \\
K 2-17   &   336.8   &   -7.2   &   139   &   13   &   DSS2    \\
G337.0+08.4   &   337   &   8.4   &   0   &   41   &   MASH    \\
G337.3+00.6   &   337.3   &   0.6   &   124   &   171   &   MASH    \\
G337.8-04.1   &   337.8   &   -4.1   &   50   &   101   &   MASH    \\
G338.0+02.4   &   338   &   2.4   &   130   &   176   &   MASH    \\
NGC 6326   &   338.1   &   -8.3   &   54   &   110   &   DSS2    \\
He 2-155   &   338.8   &   5.6   &   61   &   106   &   DSS2    \\
G338.9+04.6   &   338.9   &   4.6   &   30   &   75   &   MASH    \\
G339.1+00.9   &   339.1   &   0.9   &   164   &   32   &   MASH    \\
G339.4-06.5   &   339.4   &   -6.5   &   13   &   67   &   MASH    \\
G340.0+02.9   &   340   &   2.9   &   0   &   47   &   MASH    \\
Sa 1-6   &   340.4   &   -14.1   &   42   &   106   &   DSS2    \\
Lo 11   &   340.8   &   12.3   &   61   &   104   &   DSS2    \\
Lo 12   &   340.8   &   10.8   &   86   &   130   &   DSS2    \\
G340.9+03.7   &   340.9   &   3.7   &   141   &   8   &   MASH    \\
NGC 6026   &   341.6   &   13.7   &   63   &   106   &   DSS2    \\
G341.7+02.6   &   341.7   &   2.6   &   68   &   116   &   MASH    \\
G342.0-01.7   &   342   &   -1.7   &   90   &   142   &   MASH    \\
NGC 6072   &   342.1   &   10.8   &   71   &   116   &   DSS2    \\
Sp 3   &   342.5   &   -14.3   &   144   &   28   &   DSS2    \\
H 1-3   &   342.7   &   0.7   &   146   &   17   &   DSS2    \\
He 2-207   &   342.9   &   -4.9   &   56   &   112   &   DSS2    \\
Pe 1-8   &   342.9   &   -2   &   61   &   114   &   DSS2    \\
SuWt 3   &   343.6   &   3.7   &   77   &   127   &   DSS2    \\
G343.9-01.6   &   343.9   &   -1.6   &   116   &   169   &   MASH    \\
H 1-6   &   344.2   &   -1.2   &   145   &   18   &   DSS2    \\
H 1-7   &   345.2   &   -1.2   &   66   &   120   &   CAM98    \\
Tc 1            &   345.2   &   -8.8   &   R   &   ...   &   HST    \\
MeWe 1-11   &   345.3   &   -10.2   &   41   &   102   &   DSS2    \\
IC 4637   &   345.4   &   0.1   &   0   &   53   &   DSS2    \\
He 2-175   &   345.6   &   6.7   &   94   &   143   &   DSS2    \\
G345.8+02.7   &   345.8   &   2.7   &   58   &   109   &   MASH    \\
Vd 1-6   &   345.9   &   3   &   110   &   162   &   DSS2    \\
IC 4663   &   346.2   &   -8.2   &   86   &   146   &   DSS2    \\
A 38   &   346.9   &   12.4   &   83   &   130   &   DSS2    \\
G347.0+00.3   &   347   &   0.3   &   58   &   111   &   MASH    \\
G347.2-00.8   &   347.2   &   -0.8   &   97   &   151   &   MASH    \\
G347.4+01.8   &   347.4   &   1.8   &   161   &   33   &   MASH    \\
IC 4699   &   348   &   -13   &   28   &   92   &   CAM98    \\
G349.1-01.7   &   349.1   &   -1.7   &   114   &   169   &   MASH    \\
NGC 6337   &   349.3   &   -1.1   &   139   &   15   &   DSS2    \\
NGC 6302   &   349.5   &   1.1   &   83   &   137   &   DSS2    \\
H 1-26   &   350.1   &   -3.9   &   88   &   146   &   DSS2    \\
G350.9-02.9   &   350.9   &   -2.9   &   53   &   110   &   MASH    \\
H 2-1          &   350.9   &   4.4   &   7   &   59   &   HST    \\
G351.1-03.9   &   351.1   &   -3.9   &   170   &   47   &   MASH    \\
M 1-19          &   351.2   &   4.8   &   0   &   52   &   HST    \\
G352.2+02.4   &   352.2   &   2.4   &   96   &   150   &   MASH    \\
K 2-16   &   352.9   &   11.4   &   141   &   12   &   DSS2    \\
G353.3-02.9   &   353.3   &   -2.9   &   55   &   113   &   MASH    \\
Wray 16-411   &   353.7   &   -12.8   &   87   &   152   &   DSS2    \\
G353.8-03.7   &   353.8   &   -3.7   &   177   &   55   &   MASH    \\
G353.9-05.8   &   353.9   &   -5.8   &   36   &   96   &   MASH    \\
G354.0+04.7   &   354   &   4.7   &   132   &   6   &   MASH    \\
G354.0-00.8   &   354   &   -0.8   &   108   &   165   &   MASH    \\
G354.5+04.8   &   354.5   &   4.8   &   112   &   166   &   MASH    \\
G354.5-03.9   &   354.5   &   -3.9   &   31   &   90   &   MASH    \\
G354.9-04.4   &   354.9   &   -4.4   &   41   &   100   &   MASH    \\
G355.0+05.8   &   355   &   5.8   &   25   &   79   &   MASH    \\
G355.1+03.7   &   355.1   &   3.7   &   10   &   65   &   MASH    \\
G355.3+05.2   &   355.3   &   5.2   &   10   &   64   &   MASH    \\
G355.3-04.1   &   355.3   &   -4.1   &   130   &   9   &   MASH    \\
Hf 2-1   &   355.4   &   -4   &   40   &   99   &   CAM98    \\
M 3-14   &   355.4   &   -2.4   &   164   &   43   &   DSS2    \\
G355.4+03.6   &   355.4   &   3.6   &   167   &   42   &   MASH    \\
G355.6+04.1   &   355.6   &   4.1   &   24   &   79   &   MASH    \\
G355.6-02.3   &   355.6   &   -2.3   &   50   &   108   &   MASH    \\
G355.9+04.1   &   355.9   &   4.1   &   45   &   100   &   MASH    \\
G355.9-04.4   &   355.9   &   -4.4   &   131   &   10   &   MASH    \\
M 1-30          &   355.9   &   -4.3   &   98   &   157   &   HST    \\
H 1-9          &   356   &   3.6   &   ...   &   ...   &   HST    \\
H 2-26          &   356.1   &   -3.4   &   132   &   11   &   HST    \\
Th 3-3   &   356.5   &   5.1   &   102   &   157   &   DSS2    \\
G356.5+02.2   &   356.5   &   2.2   &   65   &   121   &   MASH    \\
G356.6+02.3   &   356.6   &   2.3   &   33   &   89   &   MASH    \\
G356.6-01.9   &   356.6   &   -1.9   &   153   &   31   &   MASH    \\
H 1-39          &   356.6   &   -4   &   31   &   90   &   HST    \\
M 2-24   &   356.9   &   -5.8   &   62   &   123   &   DSS2    \\
M 3-7          &   357.1   &   3.6   &   ...   &   ...   &   HST    \\
G357.2+07.2   &   357.2   &   7.2   &   55   &   109   &   MASH    \\
G357.3+01.3   &   357.3   &   1.3   &   50   &   107   &   MASH    \\
TrBr 4   &   357.6   &   1   &   26   &   84   &   DSS2    \\
G357.6-03.0   &   357.6   &   -3   &   134   &   13   &   MASH    \\
G357.6-06.5   &   357.6   &   -6.5   &   37   &   98   &   MASH    \\
G357.8+01.6   &   357.8   &   1.6   &   155   &   32   &   MASH    \\
Bl D   &   358.2   &   -1.1   &   149   &   28   &   DSS2    \\
M 3-39   &   358.5   &   5.4   &   35   &   91   &   DSS2    \\
NGC 6563   &   358.5   &   -7.3   &   60   &   121   &   DSS2    \\
G358.8-07.6   &   358.8   &   -7.6   &   34   &   96   &   MASH    \\
M 3-9   &   359   &   5.1   &   116   &   173   &   DSS2    \\
M 1-26          &   359   &   -0.7   &   ...   &   ...   &   HST    \\
A 40   &   359.1   &   15.1   &   72   &   125   &   DSS2    \\
19w32   &   359.2   &   1.2   &   50   &   108   &   CAM98    \\
Hb 5   &   359.3   &   -0.9   &   76   &   135   &   DSS2    \\
G359.3-02.3   &   359.3   &   -2.3   &   4   &   63   &   MASH    \\
G359.7-04.4a   &   359.7   &   -4.4   &   0   &   60   &   MASH    \\
\enddata
\end{deluxetable}

\clearpage


\clearpage

\clearpage


\clearpage
\clearpage


\clearpage
\clearpage


\clearpage
\clearpage


\clearpage


\clearpage

\clearpage \clearpage

\label{lastpage}

\end{document}